\documentstyle[aps,epsf,prl, multicol]{revtex}
\begin{document}
\renewcommand{\thesection}{\Roman{section}}
 \title{Macroscopic Quantum Self-trapping and Atomic Tunneling 
in  Two-species Bose-Einstein Condensates} 
\author{Le-Man Kuang$^{1,2}$,  Zhong-Wen Ouyang$^2$}
\address{$^1$CCAST (World Laboratory), P.O. Box 8730, Beijing 100080,
             China\\
         $^2$Department of Physics, Hunan Normal University, Changsha 410081,
             China$^*$}
\date{\today }
\maketitle
\begin{abstract}
We present a new theoretical treatment of macroscopic quantum
self-trapping (MQST) and quantum coherent atomic tunneling   in a zero-temperature
two-species Bose-Einstein condensate system in the presence of the
nonlinear self-interaction of each species, the interspecies nonlinear interaction, and the
Josephson-like tunneling interaction. It is shown that the  nonlinear
interactions can  dramatically affect the MQST and the atomic tunneling, and lead to the collapses
and revivals (CR) of population imbalance between the two condensates. The competing effects
between the self-interaction of each species and the interspecies
interaction can lead to the quenching of the MQST and the suppression  of the CR
and the Shapiro-like steps of the atomic tunneling  current. It is revealed that the
interatomic nonlinear interactions  can induce the  coherent atomic tunneling  between
 two condensates  even though there does not exist the interspecies
 Josephson-like tunneling  coupling.

\noindent PACS number(s): 03.75.Fi, 74.50.+r, 05.30.Jp, 32.80.Pj
\end{abstract}

\begin{multicols}{2}
\section{Introduction}
Recently,  much attention has been paid to the investigations  of
systems consisting of two weakly interacting Bose-Einstein condensates [1-17]  due to
the appearance  of  quantum interference [18-34] and  new
macroscopic quantum phenomena [35-37]. In principle, such condensate systems can be
 produced in a double trap with two condensates coupled by quantum
 tunneling and ground collisions, or in a system with two different
 magnetic sublevels of an atom, in which case the two species
 condensates correspond two electronic states involved. The  coupling
 between two condensates could be realized by the near-resonant
 dipole-dipole  interaction.
 
   The first  experiment [6] involving interactions between two
 condensates was performed with atoms evaporatively cooled in the
  $|F=2, M_f=2\rangle $ and $|1,-1\rangle $ spin states of $^{87}Rb$. One of the
 latest experimental advances [3,7] in this direction is the realization of  measurements 
 of relative phase in two-component Bose-Einstein condensates. In experiments at JILA [3], 
 two condensates in two different internal atomic states are produced by using a single two-photon 
 coupling pulse.   The two condensates have a well defined relative phase. After a time  during which 
  the condensates evolve in the trapping potentials, the two condensates interfere through  mixing 
   coherently the two internal   atomic states. Then, the relative phase of 
   the two condensates is obtained from the spatial interference pattern.
    The realization of  measurements of the relative phase between two condensates opened 
    the fascinating possibility   of experimentally examing  the phase-related phenomena
    in Bose condensates,  such as atomic Josephson effect and 
 macroscopic quantum self-trapping (MQST) [37].
 
 Theoretical studies of such systems began in Ho and Shenoy's work
[9] which shown that binary mixtures of condensates of alkali atoms have
a great variety of ground state and vortex structures. Then, the
stability and collective excitations of two-species condensate systems
[1,2,4,5] have been extensively studied.
More recently, Smerzi and coworkers [37] have shown that the in a  system  of two Bose condensates 
the quantum  coherent atomic tunneling  between two condensates 
induces two types of interesting effects. One is an atomic Josephson effect in Bose condensates,  which [43]
 is a generalization of the sinusoidal  Josephson effects familiar in superconductors. 
  The other is   macroscopic quantum self-trapping (MQST), which is a kind of a self-locked 
  population imbalance between  two Bose condensates.  
 It  arises because of the interatomic  nonlinear self-interaction.  The MQST has a quantum nature, 
 involving  the coherence of a macroscopic number of atoms in the two condensates.  
 It has been known that the MQST depends upon the trap parameters, the total atoms and initial states 
 of the system  and is self-maintained in a closed     conserved  system without external drives.  
  As pointed out in Ref.[37],  it is easier to observe the MQST in Bose condensates than 
   self-trapping phenomena   in other systems,  such as the  single-electron  Coulomb Blockade 
   effect [38]    arising from the Coulomb interaction 
  between electrons,  single polaron trapping in a medium [39] which arises from single  electrons, 
  interacting with a polarizable lattice, and external gravitational effects on He II baths [40,41].
  
     In Ref.[37], Smerzi and coworkers only considered the MQST induced by interatomic nonlinear self-interaction 
     in each condensate.     However, in a system  consisting of two Bose condensates, there are not only  
    nonlinear self-interaction  but also  interspecies nonlinear interaction.    
     Questions that naturally arise are, what is the effect of the interspecies nonlinear interaction on the MQST 
     and quantum coherent atomic tunneling?  Does interspecies  nonlinear interaction strengthen or weaken the MQST 
     and  the atomic tunneling  current between them?  
 
 In this paper, we present a theoretical treatment of the MQST and the
quantum coherent atomic tunneling  in a  more general two-species Bose condensate system in terms of
a two-mode approximate model and the rotating wave approximation. Our treatment involves not only 
the  interatomic nonlinear self-interaction in each species but also the interspecies nonlinear interaction. 
We find that the presence of the interspecies nonlinear interaction gives rise to new insight to 
    the MQST and the atomic tunneling  between the two condensates.
This paper is organized as follows. In Sec. II, we establish our model and present an approximate analytic solution. 
In Sec. III, we discuss the collapse and revival (CR) phenomenon on population imbalance between two condensates. 
In Sec. IV, we investigate the MQST in the two-condensate system, and discuss the dependence of the MQST 
upon  the initial states and the tunneling interaction and the nonlinear interactions. In Sec. V, we study quantum 
dynamics  of the atomic tunneling current and its $dc$ characteristics. We shall conclude our paper with discussions 
and remarks in the last section.

\section{Model and Solution}

We consider a zero-temperature two-species Bose  condensate  system in which the atoms
 interact via $aa$ and $bb$ and $ab$ elastic collisions, and  there is a Josephson-like
coupling term denoted by $a^{\dagger}b$ and $ab^{\dagger}$. In the formalism of
 the second quantization, Hamiltonian of  such a system can be written as
\begin{eqnarray}
\hat{H}&=&\hat{H}_1+\hat{H}_2+\hat{H}_{int} +\hat{H}_{Jos} , \\
\hat{H}_i&=&\int d{\bf x} \hat{\psi}^{\dagger}_i({\bf x})[-\frac{\hbar^2}{2m}\nabla^2 +
V_i({\bf x}) \nonumber \\
&&+ U_i\hat{\psi}^{\dagger}_i({\bf x})\hat{\psi}_i({\bf x})]\hat{\psi}_i({\bf x}),
\hspace{0.5cm}(i=1,2),\\
\hat{H}_{int}&=&U_{12} \int d{\bf x}
\hat{\psi}^{\dagger}_1({\bf x})\hat{\psi}^{\dagger}_2({\bf x})\hat{\psi}_1({\bf x})\hat{\psi}_2({\bf x}), \\
\hat{H}_{Jos}&=&\Lambda \int d{\bf x}
[\hat{\psi}^{\dagger}_1({\bf x})\hat{\psi}_2({\bf x}) + \hat{\psi}_1({\bf x})\hat{\psi}^{\dagger}_2({\bf x})],
\end{eqnarray}
where $\hat{\psi}_i({\bf x})$ and $\hat{\psi}^{\dagger}_i({\bf x})$ are the atomic field
operators which annihilate and create atoms
at position ${\bf x}$, respectively, they  satisfy   the  commutation
relation 
\begin{equation}
[\hat{\psi}_i({\bf x}), \hat{\psi}^{\dagger}_j({\bf x}')]=\delta_{ij}\delta({\bf x}-{\bf x}').
\end{equation}
In Eq.(1), $\hat{H}_1$ and $\hat{H}_2$  describe the evolution of each condensate  in the absence of
interspecies interaction. $\hat{H}_{int}$ describes interspecies collisions. $\hat{H}_{Jos}$ is the 
Josephson-like tunneling coupling term.  Atoms are confined in harmonic potentials 
$V_i({\bf x}) (i=1,2)$ of frequencies
 $\omega_i$. Interactions between atoms are described by a nonlinear  self-interaction term
 $U_i=4\pi\hbar^2a^{sc}_i/m$ and a term that corresponds the nonlinear interaction
 between different condensates  $U_{12}=4\pi\hbar^2a^{sc}_{12}/m$, where    $a^{sc}_i$
is  $s$-wave scattering length of condensate  $i$ and  $a^{sc}_{12}$ that between condensate  1 and 2.
 For simplicity, throughout this paper we let $\hbar=1$ and  assume that $a^{sc}_1=a^{sc}_2=a^{sc}$, and 
$V_1({\bf x})=V_2({\bf x})$.

It is well known that  the above Hamiltonian can be reduced to two-mode boson Hamiltonian [30-32, 35,36] through 
expanding  the atomic field operators over  single-particle states [35]:
\begin{equation}
\hat{\psi}_i({\bf x})=\hat{a}_i\phi_{iN}({\bf x}) + \tilde{\psi}_i({\bf x}),
\end{equation}
where $\hat{a}^{\dagger}_i=\int d{\bf x}\phi_{iN}({\bf x})\hat{\psi}^{\dagger}_i({\bf x})$
create particles with distributions $\phi_{iN}({\bf x})$ with $[\hat{a}_i, \hat{a}^{\dagger}_i]=1$.
  The first term in the  mode expansion (6)   acts  only on the condensate state vector, whereas the second term
 $\tilde{\psi}_i({\bf x})$   accounts for  noncondensed  atoms.
Substituting the mode expansions of the atomic field operators  into the Hamiltonian (1), retaining only the
 first term representing the condensates, we arrive at  the following
 two-mode approximate Hamiltonian 
\begin{eqnarray}
 \hat{H} &=&\omega_0(\hat{a}^{\dagger}_1\hat{a}_1 + \hat{a}^{\dagger}_2\hat{a}_2)
+  q(\hat{a}^{\dagger 2}_1\hat{a}^2_1 + \hat{a}^{\dagger 2}_2\hat{a}^2_2)\nonumber \\
&&+  g(\hat{a}^{\dagger}_1\hat{a}_2 + \hat{a}^{\dagger}_2\hat{a}_1) + 2\chi\hat{a}^{\dagger}_1\hat{a}_1
\hat{a}^{\dagger}_2\hat{a}_2,
\end{eqnarray}
where the frequency and the coupling constants are defined by 
\begin{eqnarray}
 \omega_0&=&\sum^{2}_{i=1}\int d{\bf x}
[\frac{1}{2m}|\nabla\phi_{iN}({\bf x})|^2+V({\bf x})(|\phi_{iN}({\bf x})|^2], \\
  q&=&U_0\int d{\bf x} (|\phi_{1N}({\bf x})|^2 + |\phi_{2N}({\bf x})|^2), \\
  g&=&\Lambda\int d{\bf x} (\phi_{1N}^{\dagger}({\bf x})\phi_{2N}({\bf x}) +
\phi_{1N}({\bf x})\phi_{2N}^{\dagger}({\bf x})),\\
  \chi&=&\frac{1}{2}U_{12}\int d{\bf x}
|\phi_{1N}^{\dagger}({\bf x})|^2|\phi_{2N}({\bf x})|^2.
\end{eqnarray}

From Eqs.(6) and (7) we can see that the two-mode approximation essentially consists in neglecting all modes 
except the condensate modes. At zero temperature, this amounts to ignoring the atoms which have left the 
condensate mode  due to the two-body interactions.  In other words, what the two-mode approximation involves  
is only   the first order effects of interactions.  The mode expansion of the condensate function over 
single-particle states (6) makes the condensate 
shape not to be changed, this limits migration of  condensed  atoms from one condensate to the other. 
The constraint on the shapes of  condensates    implies that  the two-mode approximation can be applied 
only for weak nonlinearity. The valid conditions of the two-mode approximation were demonstrated in Refs.[30, 32,35], 
which indicate that this approximation provides a reasonably accurate picture for weak many-body interactions , i.e., 
for   small number of   condensed  atoms.  For large condensates, the mode functions of condensates are altered 
due to the   collisional  interactions, and the two-mode approximation breaks down.  As shown in Ref.[30,32],  a simple 
estimate shows that this  happens when the number of atoms $Na^{sc}\gg r_0$, where $a^{sc}$ is a typical scattering length 
and $r_0$ is  a measure of the trap size. If we consider a large trap with the size $r_0=10 \mu$m and 
 the  typical scattering length $a^{sc}=5$ nm, the two-mode approximation is applicable for  $N\le 2000$. 
 This is the case which we consider here. We shall show that the MQST and atomic tunneling 
 between the two condensates  are strongly affected   by the nonlinear many-body interactions.

We note that the  two-mode approximate Hamiltonian has the same form with  that of a two-mode
nonlinear optical directional coupler [42]. The two-mode Hamiltonian (7)
can not be exactly solved, but for weak nonlinear interactions a closed analytical solution can be
obtained under the rotating wave approximation suggested by Alodjanc
{\it et al.} [43].  

In order to obtain  an approximate analytic solution of the Hamiltonian
(7), we introduce a new  pair of  bosonic operators $\hat{A}_1$ and $\hat{A}_2$ by the following expressions::
\begin{equation}
\hat{a}_1=\frac{1}{\sqrt{2}}(\hat{A}_1e^{igt} -i\hat{A}_2e^{-igt}),\\
\hat{a}_2=\frac{1}{\sqrt{2}}(\hat{A}_1e^{igt} +i\hat{A}_2e^{-igt}),
\end{equation}
 where $\hat{A}_1$ and $\hat{A}_2$ are slowly varying operators, they  satisfy the usual bosonic commutation relations:
 $[\hat{A}_i, \hat{A}_j]=0$, and   $[\hat{A}_i, \hat{A}^{\dagger}_j]=\delta_{ij}$ with $\hat{A}^{\dagger}_i$ being 
 the hermitain conjugation of $\hat{A}_j$. Then the Hamiltonian  (7) reduces to the following form
\begin{eqnarray}
  \hat{H} &=&\omega\hat{N}+\frac{1}{4}q[3\hat{N}^2 -(\hat{A}^{\dagger}_1\hat{A}_1-
 \hat{A}^{\dagger}_2\hat{A}_2)^2] \nonumber \\
 &&+ g(\hat{A}^{\dagger}_1\hat{A}_1-
 \hat{A}^{\dagger}_2\hat{A}_2)+\frac{\chi}{2}\hat{N}^2 \nonumber  \\
 &&-\chi\hat{A}^{\dagger}_1\hat{A}_1\hat{A}^{\dagger}_2\hat{A}_2+\hat{H}',
\end{eqnarray} 
 where the detuning is given by  $\omega=\omega_0- (\chi+q)/2$, the total number operator $\hat{N}$ is  
 a   conserved  constant which  is given by
\begin{equation}
\hat{N}=\hat{a}^{\dagger}_1\hat{a}_1+\hat{a}^{\dagger}_2\hat{a}_2=\hat{A}^{\dagger}_1\hat{A}_1+\hat{A}^{\dagger}_2\hat{A}_2,
\end{equation}
and $\hat{H}'$ is a  nonresonant term which is given by 
\begin{equation}
\hat{H}'_1=\frac{1}{2}(\chi-q)(\hat{A}^{\dagger 2}_1\hat{A}^2_2 e^{-i4gt} + 
\hat{A}^{\dagger 2}_2\hat{A}^2_1 e^{i4gt}), 
\end{equation}
  which oscillates   at the frequency $4g$. The account of the fast oscillating term  results only in some 
 addtional oscillations which play no essential role in the evolution of the measurable quantities  
 specifying the macroscopic quantum phenomena of the two-condensate system, so that it  is fully  negligible. 
  This means  the  rotating wave approximation [43].
  After neglecting the nonresonant term   $H'$,  we get the following approximate Hamiltonian:
\begin{eqnarray}
\hspace{-1.0cm}
 \hat{H}_A &=&\omega\hat{N} + g(\hat{A}^{\dagger}_1\hat{A}_1-\hat{A}^{\dagger}_2\hat{A}_2)
 \nonumber \\&&+\frac{1}{4}q[3\hat{N}^2 -(\hat{A}^{\dagger}_1\hat{A}_1-\hat{A}^{\dagger}_2\hat{A}_2)^2]
 \nonumber \\ &&+\frac{1}{2}\chi\hat{N}^2-\chi\hat{A}^{\dagger}_1\hat{A}_1\hat{A}^{\dagger}_2\hat{A}_2.
\end{eqnarray}

In order to solve the Hamiltonian (16) we introduce  two Fock spaces  of 
 $(\hat{A}_1, \hat{A}_2)$ and $(\hat{a}_1, \hat{a}_2)$ in which the bases are defined by 
\begin{eqnarray}
 |n,m)&=&\frac{1}{\sqrt{n!m!}}\hat{A}^{\dagger n}_1\hat{A}^{\dagger m}_2|0,0), \\ 
 |n,m\rangle &=&\frac{1}{\sqrt{n!m!}}\hat{a}^{\dagger n}_1\hat{a}^{\dagger m}_2|0,0\rangle .
\end{eqnarray}
where $n$ and $m$ take non-negative integers.  Obviously, $\hat{H}_A$ is diagonal in the Fock space
of  $(\hat{A}_1, \hat{A}_2)$, and we find that  
 \begin{eqnarray}
\hat{H}_A|n,m)&=&E(n,m)|n,m), \\
 E(n,m)&=& \omega (n+m)+g(n-m)
 \nonumber  \\&&
 +\frac{1}{4}(3q+\chi)(n+m)^2-\frac{1}{4}q(n-m)^2
 \nonumber  \\&&-\chi nm.
\end{eqnarray}
 
 Consider  two coherent states defined in Fock spaces of $(\hat{A}_1, \hat{A}_2)$ and $(\hat{a}_1, \hat{a}_2)$, 
 respectively,
 \begin{eqnarray}
 |\alpha_1, \alpha_2\rangle &=&D_{\hat{a}_1}(\alpha_1)D_{\hat{a}_2}(\alpha_2)|0,0\rangle ,
 \\
|u_1,u_2) &=&D_{\hat{A}_1}(u_1)D_{\hat{A}_2}(u_2)|0,0),
\end{eqnarray}
 where $D_{\hat{a}_i}(\alpha_i)$ and  $D_{\hat{A}_i}(u_i)$ are displacement 
 operators    defined by  
\begin{eqnarray}
D_{\hat{a}_i}(\alpha_i)&=&\exp(\alpha_i\hat{a}_i+\alpha^*_i\hat{a}^{\dagger}_i), \\
D_{\hat{A}_i}(u_i)&=&\exp(u_i\hat{A}_i+u^*_i\hat{A}^{\dagger}_i).
\end{eqnarray}
  
   Note the fact that $|0,0)=|0,0\rangle $, we can find a useful relation to connect
$|\alpha_1, \alpha_2\rangle $ and  $|u_1,u_2)$ with each other
\begin{eqnarray}
|\alpha_1, \alpha_2\rangle &=&|\frac{\alpha_1 + \alpha_2}{\sqrt{2}}, \frac{i(\alpha_1 -
\alpha_2)}{\sqrt{2}}), \\
|\alpha_1, \alpha_2)&=&|\frac{\alpha_1 - i\alpha_2}{\sqrt{2}}, \frac{\alpha_1 +
i\alpha_2}{\sqrt{2}}\rangle .
\end{eqnarray}

Following  the arguments of Bose broken symmetry, we assume that
 the two condensates are initially in the coherent states $|\alpha_1\rangle $ and
 $|\alpha_2\rangle $,  which are eigenstates of $\hat{a}_1$ and $\hat{a}_2$,
 respectively. Then the wave function of the two species condensate system
 at time $t$ can be explicitly expressed as
\begin{eqnarray}
|\Phi(t)\rangle &=&e^{-\frac{1}{2}N}\sum^{\infty}_{n,m=0}\frac{1}{\sqrt{n!m!}}
u_1^n(iu_2)^m \nonumber \\
&&\times e^{-iE(n,m)t}|n,m),
\end{eqnarray}
where
\begin{eqnarray}
u_1&=&\frac{1}{\sqrt2}(\alpha_1+\alpha_2),  u_2=\frac{1}{\sqrt2}(\alpha_1-\alpha_2) \\
N&=&|\alpha_1|^2+|\alpha_2|^2=|u_1|^2+|u_2|^2, 
\end{eqnarray}
where we have used Eqs.(25), (26) in the derivation of Eq.(27).

\section{collapse  and revivals of Population imbalance}

In this section we show that   the two condensate system under our consideration 
exhibits  a  collapse and revival phenomenon of population imbalance between two condensates. 
Denote  the number difference  of atoms between the two condensates by 
\begin{equation}
D(t)= N_1(t)-N_2(t).
\end{equation}
Then from Eq.(27) we can find that at time $t$, the   number of  atoms in each 
 condensate  
$N_i(t)=\langle \hat{a}^{\dagger}_i\hat{a}_i\rangle$  is given by
\begin{eqnarray}
N_i(t)&=&\frac{1}{2}\{N-(-1)^i2|u_1||u_2|\cos[4gt+\theta(t)]
\nonumber \\
&&\times e^{-2N\sin^2\frac{1}{2}(q-\chi)t}\},  (i=1,2)
\end{eqnarray}
where we have used the following  symbols:
\begin{equation}
u_i=|u_i|e^{i\varphi_{u_i}}, \\ 
\theta(t)= (\varphi_{u_2}-\varphi_{u_1})+u_{21}\sin(q-\chi)t,
\end{equation}
 with $u_{21}$ and $\varphi_{u_i}$  being defined by 
\begin{eqnarray}
u_{21}&=&|u_2|^2-|u_1|^2,  \\
\varphi_{u_i}&=&\tan^{-1}\{\frac{|\alpha_1|\sin\varphi_1\mp(-1)^{i}|\alpha_2|\sin\varphi_2}
                        {|\alpha_1|\cos\varphi_1\mp(-1)^{i}|\alpha_2|\cos\varphi_2}\}.
\end{eqnarray}
Then, the  population difference  is given by
\begin{eqnarray}
 D(t)&=&2|u_1||u_2|\cos[4gt+ \theta(t)]
\nonumber \\ 
&&\times e^{-2N\sin^2(\frac{1}{2}(q-\chi)t},
\end{eqnarray}
where $N=|\alpha_1|^2+|\alpha_2|^2$ is the total number of the atoms in the two condensates. 
Eq.(35) indicates that the population imbalance periodically oscillates with the time evolution.
From Eq.(35) we can see that $D(t)$ exhibits collapse and revival 
phenomenon which is a kind of nonclassical effect  well known in    the Jaynes-Cummings model [44] 
to describe interaction between a single-mode radiation field and a two-level atom. The CR is
also found in a Bose  condensate system [35,36].  From Eq.(35) we see that  the CR  of 
 the population imbalance in the two-condensate system depends on the  tunneling interaction ($g$) and
interatomic nonlinear  interactions ($q$ and $\chi$). When $g>|q-\chi|/8$,
since the function $\cos[4gt+\theta(t)]$  is the rapidly varying part in
(35), so that the shape of the CR
is determined by the envelope function  $\exp[-2N\sin^2(\frac{1}{2}(q-\chi)t]$.
The maximal revivals take place at time $t=2n\pi/|q-\chi|$, where $n$ is an integer.
When $g<|q-\chi|/8$, the function $\exp[-2N\sin^2(\frac{1}{2}(q-\chi)t]$ becomes
  the rapidly varying part in (35),  the CR then is determined by the envelope function
$\cos[4gt+\theta(t)]$. 
In Fig. 1  we plot the evolution of  the population  difference   between the two
condensates  with respect to the time which is in units of $|q-\chi|$, when the two condensates are in 
 the initial state of $\alpha_1= 5$ and $\alpha_2=4$, and the tunneling coupling is $g=25|q-\chi|$.  
Fig. 1 clearly indicates the CR phenomenon of  the population  difference.

It is worthwhile to  note that when $q=\chi$,  the population difference (35) becomes
\begin{equation}
D(t)=2|u_1||u_2| \cos[4gt+(\varphi_{u_2}-\varphi_{u_1})],
\end{equation}
which is a simple  sinusoidal oscillation, no CR occurs. The suppression of
the CR can be explained by looking at the expression (35) which indicates
that both the self-interaction ($q$) and the interspecies interaction
($\chi$) can induce the CR, but the CR produced by one can weaken that
by another. It is the CR produced by the self-interaction completely
counteracts the CR by the interspecies interaction that leads to the
suppression of the CR.

When there is no nonlinear interactions, i.e., $q=\chi=0$, from Eq.(35) it is easy to 
find that the population imbalance has the same form with that 
of the case $q=\chi\neq 0$. This means that the CR vanishes when the 
 nonlinearity vanishes. Hence, the CR of the population imbalance is 
 a consequence of nonlinear interactions in condensates. 
The CR of the oscillatory transfer of atoms between the two condensates constitutes a
novel macroscopic quantum phenomenon  induced by interatomic nonlinear interactions for 
the two species condensate system.

\section{Macroscopic Quantum Self-trapping}

In this section we are concerned with the MQST.
The MQST effect is characterized by the nonzero time mean value of the fractional population  
imbalance  between the two condensates defined by
\begin{equation}
p(t)=\frac{N_1(t)-N_2(t)}{N}.
\end{equation}

From Eqs.(35) and (37) we get that 
\begin{eqnarray}
 p(t)&=&\frac{2|u_1||u_2|}{N}\cos[4gt+ \theta(t)]
\nonumber \\ 
&&\times e^{-2N\sin^2(\frac{1}{2}(q-\chi))t}.
\end{eqnarray}

In order to investigate the MQST, we expand the above equation as
\begin{eqnarray}
p(t)&=&\frac{2|u_1||u_2|}{N}
e^{-N}\sum^{+\infty}_{n,m=-\infty}J_n(u_{21})I_m(N)\nonumber \\ 
&&\times \cos\{[(n+m)(q-\chi)+4g]t
\nonumber \\
&&+(\varphi_{u_2}-\varphi_{u_1})\},
\end{eqnarray}
where $J_n(A)$ and $I_n(A)$ are Bessel function and modified Bessel
function.
 From Eq.(35) it is easy to find that when the tunneling interaction 
 and nonlinear interactions satisfy the condition: 
  \begin{equation}
 4g=K(\chi-q),
 \end{equation}
 where $K$ is an integer, we can get a nonzero  time-averaged value of population imbalance
\begin{eqnarray}
\bar{p}&=&\frac{2|u_1||u_2|}{N}
e^{-N}\sum^{+\infty}_{n=-\infty}J_n(u_{21})I_{K-n}(N)
\nonumber \\
&&\times \cos(\varphi_{u_2}-\varphi_{u_1}),
\end{eqnarray}
which indicates  the existence of the MQST. Eq.(40) is the condition under which the MQST happens.

From Eq.(39) we can see that when the tunneling interaction vanishes 
and nonlinear self-interaction equals nonlinear interspecies interaction, 
i.e.,  $g=0$ and $q=\chi$, we   arrive at a constant population imbalance  
\begin{equation}
p(t)=\frac{2|u_1||u_2|}{N}\cos(\varphi_{u_2}-\varphi_{u_1}).
\end{equation}
This  is a time-independent state, called the self-trapping stationary state, 
which is the consequence of competing between nonlinear self-interaction and 
nonlinear interspecies interaction. 

When there exists the tunneling coupling, i.e., $g\neq 0$, from Eq.(40) we can 
find the critical value of the tunneling coupling at which the MQST happens
 $g_c=|q-\chi|/4$. This critical value $g_c$ depends upon  only the difference 
  between the nonlinear self-interaction and interspecies nonlinear
  interaction, not the nonlinear self-interaction and interspecies nonlinear
  interaction themselves.  Therefore, it  becomes possible that the MQST occurs
  only when the tunneling coupling  equals or exceeds the critical value $g_c$.

In Fig.2 we plot the time evolution of the fractional population imbalance when 
the two condensates are in the initial  state of  $\alpha_1= 10 $ and $\alpha_2=0$ for 
 (a) $K=1$, (b) $K=20$. Here the time  is in units of $|q-\chi|$. From Fig.2 (a) 
 and (b) we can see that the weaker 
the tunneling coupling $(g)$, the more apparent the MQST becomes. This indicates 
that the MQST is an effect induced by interatomic  nonlinear interactions $q$ and $\chi$ 
not the tunneling interaction $(g)$.  In what follows 
we shall discuss  the dependence of the MQST in detail 
upon the initial states and nonlinear interactions for specific cases.

\begin{center} 
A. {\bf  The initial state dependence}
\end{center} 

We  here  discuss the dependence of the MQST on the initial states of the
two condensates in the following four cases. 

{\it Case 1}: $|\alpha_1|=|\alpha_2|$, $\varphi_1\neq \varphi_2$.  
    In this case,   the two condensates initially have the same  number of atoms
    but different phases.  From Eq.(32) and (33)  we get that 
    $|u_1|^2=N\cos^2[(\varphi_1-\varphi_2)/2]$, $|u_2|^2=N\sin^2[(\varphi_1-\varphi_2)/2]$, 
    and $u_{21}=-N\cos(\varphi_1-\varphi_2)$.
    Making use of Eq.(28), from Eqs.(35) and  we can see that 
 the fractional population periodically evolves with respect to  time, $p(t)\neq 0$ except that
$\varphi_1-\varphi_2=n\pi$, where $n$ is an integer. If  $4g/(\chi-q)=K$
(an integer), we get the locked population imbalance 
\begin{eqnarray}
\bar{p}&=&|\sin(\varphi_1-\varphi_2)|\cos(\varphi_{u_1}-\varphi_{u_2})e^{-N}
\nonumber \\
& &\times \sum^{+\infty}_{n=-\infty}J_n(-N\cos(\varphi_1-\varphi_2))I_{K-n}(N),
\end{eqnarray}
 which implies that the MQST does exist, even if  the two condensates initially have the same 
  number of the   atoms provided that they have different initial phases.
 And  less the total atomic number is, stronger  the MQST becomes.

{\it Case 2}:   $|\alpha_1|\neq |\alpha_2|$, $\varphi_1= \varphi_2$.
    In this case  the two condensates initially  have the same phases
but the different number of atoms. From Eq.(38) we see that when    $4g/(\chi-q)=K$ 
(an integer), the MQST occurs with the following locked population  imbalance:
\begin{equation}
\bar{p}=\frac{|\alpha_1|^2+|\alpha_2|^2}{2N}e^{-N}
\sum^{+\infty}_{n=-\infty}J_n(-2|\alpha_1|\alpha_2|)I_{K-n}(N).
\end{equation}

{\it Case 3}: $N=|\alpha_1|^2$,  $|\alpha_2|=0$. 
  In this case, the system  starts with all atoms being in one condensate.
  Making use of Eq.(28), we get that $u_1=u_2=\alpha_1/\sqrt2$, and $\theta(t)=0$.  
  The fractional population evolution is given by 
\begin{equation} 
 p(t)=e^{-N}\sum^{+\infty}_{n=-\infty}I_n(N)\cos\{[n(q-\chi)+4g]t,
\end{equation}
So that  when the tunneling interaction and the nonlinear  interactions satisfy the relation 
 $4g/(\chi-q)=K$ (an integer),  we can see the appearance of the MQST with the 
 locked   population imbalance
 \begin{equation}
 \bar{p}=e^{-N}I_K(N).
 \end{equation}

{\it Case 4}:  $|\alpha_1|=|\alpha_2|$ and $\varphi_1= \varphi_2$. In this case
   the two condensates initially have the same number of atoms  and  the same phases. 
   From Eq.(28)  we have $u_2=0$. Making use of Eq.(38) we can see that the 
 oscillations of the population imbalance  vanish, i.e., $p(t)=0$, and no  MQST occurs. 
 This is in agreement with the result in Ref.[37].

\begin{center} 
B. {\bf  The dependence  on   nonlinear interactions }
\end{center} 

Then, we turn to the dependence of the MQST upon the tunneling coupling ($g$) and
the nonlinear interactions  between atoms, which  are described by
the  parameters $q$ and $\chi$
corresponding to self-interactions and interspecies interactions,
respectively.

 {\it Case 1}:   $g=0$,  $q\neq 0$, and $\chi\neq 0$.  In this case there is no tunneling  interaction, 
but there exists interatomic nonlinear interactions. From Eq.(39) we find that  
the fractional population imbalance becomes
\begin{eqnarray}
\hspace{-0.5cm}
p(t)&=&\frac{2|u_1||u_2|}{N}
e^{-N}\sum^{+\infty}_{n,m=-\infty}J_n(u_{21})I_m(N)\nonumber \\ 
&&\times \cos\{[(n+m)(q-\chi)]t+(\varphi_{u_2}-\varphi_{u_1})\},
\end{eqnarray}
which indicates that  no MQST occurs if the coupling of interatomic  self-interaction 
does not equals that of interspecies nonlinear interaction, i.e., $q\neq \chi$.
 However, when the self-interaction equals the interspecies interaction, i.e., 
  $q=\chi$, we can observe the self-trapping stationary state with 
  a constant population imbalance $p(t)=p(0)$.

{\it Case 2}:   $g\neq 0$,  $q=\chi=0$. 
 In this case, we consider only the effect of the tunneling coupling while the interatomic
 nonlinear interactions are not involved. Eq.(38) tells us that 
   the MQST  does not occur, although there exists  oscillations of the 
   population imbalance between the two condensates. This further confirms the validity of 
    Smerzi {\it et al.}'s conclusion [37] which the MQST arises from the interatomic 
    nonlinear interaction.
    
{\it Case 3}: $g\neq 0$, and $q=\chi\neq 0$. In this case there exist both the tunneling interaction 
 and the  nonlinear interactions, but self-interaction equals  interspecies interaction. 
 Frme Eq.(38) we can find  that  the population imbalance exhibits a simple oscillation with
\begin{equation} 
p(t)=\frac{2|u_1||u_2|}{N}e^{-N}\cos[4gt+(\varphi_{u_2}-\varphi_{u_1})], 
\end{equation} 
which means that  the MQST vanishes.

{\it Case 4}: $g\neq 0$,  $q\neq 0$, $\chi=0$, or $g\neq 0$,  $q= 0$, $\chi\neq
0$. In this  case, there exist the tunneling interaction and one of  the self-interaction and 
the interspecies interaction. It is easy to see that the fractional population imbalance 
 Eq.(38) reduces to 
\begin{eqnarray} 
p(t)&=&\frac{|u_1||u_2|}{N}
e^{-N}\sum^{+\infty}_{n,m=-\infty}J_n(u_{21})I_m(N)\nonumber \\
&&\times \cos\{[(n+m)\kappa+4g]t+(\varphi_{u_2}-\varphi_{u_1})\},
\end{eqnarray}
where $\kappa=q$ or $-\chi$. Eq.(49) reflects the fact that when $4g/\kappa=n+m=K$ (an integer),  
  the MQST  happens with  the nonzero $\bar{p}$ given by Eq.(41). This
implies that both the nonlinear  self-interaction in each condensate and the
interspecies nonlinear interactions contribute to the MQST.
Since the values of $K$ to determine $\bar{p}$ have opposite signs for
the self-interaction ($\chi$) and the interspecies interaction ($q$),
the MQST produced by the self-interaction can weaken that by the
interspecies interaction. It is  the competition between the MQST induced
by the nonlinear self-interactions of each condensate and that by the
interspecies nonlinear interaction that leads to the quenching of the MQST 
in the above case 3.

\section{The Coherent Atomic Tunneling Current}

In this section, we study  quantum dynamics of the coherent 
 atomic tunneling  current between two condensates and  its $dc$ characteristics,  and discuss 
  the influence of the initial state of condensates  and the tunneling interaction   and the nonlinear 
  interactions. The coherent atomic tunneling  current between the two condensates is  defined by 
$I(t)=\dot{N}_1(t)-\dot{N}_2(t)$. Making use of Eq.(31), 
it is straightforward to get that
\begin{eqnarray}
\hspace{-0.5cm}
I(t)&=&-2|u_1||u_2|\{4g\sin(\theta(t))\nonumber \\ 
&&+(q-\chi)[|u_1|^2\sin((q-\chi)t-\theta(t))  \nonumber \\
& &+|u_2|^2\sin((q-\chi)t+\theta(t))]\}.
\end{eqnarray}
This indicates that the atomic tunneling  current periodically changes, atoms periodically transfer
between the two condensates with the time evolution. 

 In order to see $dc$ characteristic  of the atomic tunneling  current, we expand the
atomic tunneling  current (50)  as the following expression
\begin{eqnarray}
I(t)&=&-2|u_1||u_2|e^{-N} \sum^{+\infty}_{n,m=-\infty}
  \{8gJ_n(u_{21})\nonumber \\
& &-(q-\chi )[|u_1|^2J_{n-1}(u_{21})-|u_2|^2J_{n+1}(u_{21})]\}
    \nonumber \\
& &\times I_m(N)\sin\{[(m+n)(q-\chi)+4g]t
\nonumber \\
& &+(\varphi_{u_2}-\varphi_{u_1})\},
\end{eqnarray}
which implies  that when the tunneling coupling, and nonlinear couplings satisfy the condition:
 \begin{equation}
 4g=K(\chi-q),  
 \end{equation}
 we get the $dc$ component of the  atomic tunneling current with the following form,
\begin{eqnarray}
 I_{dc}(K)&=&-2|u_1||u_2|(q-\chi)\sin(\varphi_{u_2}-\varphi_{u_1})e^{-N}
\nonumber \\
& &\times \sum^{+\infty}_{n=-\infty} \{2KJ_n(u_{21})-[|u_1|^2J_{n-1}(u_{21}) 
\nonumber \\
& &-|u_2|^2J_{n+1}(u_{21})]\}I_{K-n}(N),
\end{eqnarray}
where $K$ is an integer.
This indicates that the $dc$ component of the atomic tunneling current exhibits a step structure 
with respect to the integer $K$.  This step structure   is a resonant phenomenon among  the tunneling 
interaction and nonlinear interactions with the resonant condition given by Eq.(51). It is the  analogue  
of the Shapiro steps observed in the superconductor Josephson junction [45], so that we call the steps 
 in the  step structure of the $dc$ component of the atomic tunneling current the Shapiro-like steps. 
In what follows we discuss in detail the dependence of the atomic tunneling current and the  
Shapiro-like steps upon the initial states and nonlinear interactions for some specific cases.

\begin{center} 
A. {\bf  The  initial state dependence}
\end{center}

In this subection we  discuss the initial-state dependence of the atomic tunneling  current and 
the Shapiro-like steps for  the following four cases.

{\it Case 1:} $|\alpha_1|=|\alpha_2|$, $\varphi_1\neq \varphi_2$. In this case, 
 the two condensates  initially have the same number of atoms but different phases. 
 From Eq.(50) we can get the expression of the atomic tunneling  current
\begin{eqnarray}
\hspace{-0.5cm}
I(t)&=&-N|\sin(\varphi_1- \varphi_2)|\{4g\sin(\theta(t))\nonumber \\ 
&&+(q-\chi)N[\cos^2\frac{1}{2}(\varphi_1- \varphi_2)\sin((q-\chi)t-\theta(t))  \nonumber \\
& &+ \sin^2\frac{1}{2}(\varphi_1-\varphi_2)\sin((q-\chi)t+\theta(t))]\},
\end{eqnarray}
where 
\begin{equation}
\theta(t)=(\varphi_{u_1}- \varphi_{u_2})-N\cos(\varphi_1- \varphi_2)\sin((q-\chi)t.
\end{equation}
 
 Form Eq.(54)  we can obtain the   $dc$ component of the tunneling current with the following result,
\begin{eqnarray}
 I_{dc}(K)&=&-N(q-\chi)|\sin(\varphi_1- \varphi_2)|\sin(\varphi_{u_2}-\varphi_{u_1})
\nonumber \\
& &\times e^{-N} \sum^{+\infty}_{n=-\infty} \{2KJ_n(-N\cos(\varphi_1- \varphi_2))
\nonumber \\
&&-N[\cos^2\frac{1}{2}(\varphi_1- \varphi_2)J_{n-1}( -N\cos(\varphi_1- \varphi_2) ) 
\nonumber \\
& &-\sin^2\frac{1}{2}(\varphi_1- \varphi_2)J_{n+1}( -N\cos(\varphi_1- \varphi_2))]\}
\nonumber \\
& &\times I_{K-n}(N),
\end{eqnarray}
where $n$ is an integer. Eq.(56) gives rise to the Shapiro-like steps of the atomic tunneling current.

It is interesting to note that when the initial phases of two condensates satisfy the condition:
$\varphi_1- \varphi_2=n\pi$, where $n$ is an integer, we can find zero atomic tunneling  current. This 
means that the blockade of the atomic tunneling  happens. 

When the initial phases satisfy the relation:
$\varphi_1- \varphi_2=(2n+1)\pi/2$,   $n$ being  an integer, we find that
\begin{eqnarray}
 I(t)&=&-4gN\sin(\varphi_{u_2}- \varphi_{u_1}) 
 \nonumber \\ 
&&+(q-\chi)N\cos(\varphi_{u_2}- \varphi_{u_1})\sin(2(q-\chi)t),  
\end{eqnarray}
which indicates that the atomic tunneling  current is a simple  superpositon  
of a alternating current with the sinusoidal oscillations and a $dc$ current. 
 The $dc$ component   is
 \begin{equation}
 I_{dc}=-4gN\sin(\varphi_{u_2 }- \varphi_{u_1}),
\end{equation}
which implies that the $dc$ component of the atomic tunneling  current depends only upon the tunneling 
coupling, it is independent of the nonlinear interactions in two condensates, and increases linearly 
 with  the tunneling strength $(g)$ and the the total number of the atoms $(N)$. 
 In this case no Shapiro-like steps appears.

{\it Case 2:} $|\alpha_1|\neq |\alpha_2|$,
$\varphi_1= \varphi_2$.
In this case,  the two condensates initially  have the same phases
but the  different number of atoms. The atomic tunneling current is given by  
\begin{eqnarray}
\hspace{-0.5cm}
I(t)&=&-|\alpha_1^2-\alpha_2^2|\{4g\sin(\theta(t))\nonumber \\ 
&&+\frac{1}{2}(q-\chi)[(\alpha_1+\alpha_2)^2\sin((q-\chi)t-\theta(t))  \nonumber \\
& &+ (\alpha_1-\alpha_2)^2\sin(q-\chi)t+\theta(t))]\},
\end{eqnarray}
where 
\begin{equation}
\theta(t)= -2\alpha_1\alpha_2 \sin(q-\chi)t.
\end{equation}
And the $dc$ component of the atomic tunneling  current has the following form,
\begin{eqnarray}
 I_{dc}(K)&=&- |\alpha_1^2-\alpha_2^2|(q-\chi)\sin(\varphi_{u_2}-\varphi_{u_1})
\nonumber \\
& &\times e^{-N}\sum^{+\infty}_{n=-\infty} \{2KJ_n(-2\alpha_1\alpha_2 )
\nonumber \\
& &-\frac{1}{2}[(\alpha_1-\alpha_2)^2J_{n-1}(-2\alpha_1\alpha_2 ) 
\nonumber \\
& &-(\alpha_1 -\alpha_2)^2J_{n+1}(-2\alpha_1\alpha_2)]\}I_{K-n}(N).
\end{eqnarray}

In FIG. 3, we  plot   the time evolution of  the atomic tunneling current between the two
condensates.   Results  are shown for the case of  $g=0.25$ and $q-\chi=0.1$
 when the two condensates are in the initial  state with   $\alpha_1=5$ and $\alpha_2=4$.
 FIG. 3 indicates that the atomic tunneling current exhibits complicated oscillating  behaviors.

{\it Case 3:}  $N=|\alpha_1|^2$,  $|\alpha_2|=0$. In this case,  the system   starts
 with all atoms being in one condensate. Taking into account Eqs.(32) and (33), from Eq.(50) we find 
 the atomic tunneling current to be  
\begin{equation}
 I(t)=- (q-\chi)N^2\sin(q-\chi)t,
\end{equation} 
 which is a pure sinusoidal alternating  atomic   current with the period $T=2\pi/|q-\chi|$ 
 without $dc$ component. It is worthwhile to note that   the atomic tunneling  current (62) 
  depends on only the difference between the self-coupling $q$ and the interspecies coupling $\chi$, 
  it is independent of the tunneling coupling $g$ at all. Thus, we can conclude that 
  the nonlinear interactions    can induce the atomic tunneling, even if there is no tunneling coupling 
   between two condensates.

{\it Case 4:} $|\alpha_1|=|\alpha_2|$ and $\varphi_1=
 \varphi_2$. In this  case,   the two condensates initially have the same
number of atoms and  phases. we see that the atomic tunneling  current  vanishes, and no Shapiro-like
 step occurs.

The above analyses indicate that the atomic tunneling  current and the Shapiro-like steps
strongly depend on the initial number of atoms in each  condensate  and the initial phase difference 
between the two condensates.

\begin{center} 
B. {\bf  The dependence   on nonlinear interactions }
\end{center} 

In what follows we show that the interatomic nonlinear interactions also
significantly affect the atomic tunneling  current and the Shapiro-like steps.

If we consider only the influence of  the tunneling coupling while the interatomic
nonlinear interactions are not involved, i.e., $g\neq 0$,  $q=\chi=0$, we find
that 
\begin{equation}
I(t)=-8g|u_1||u_2|\sin[4gt+(\varphi_{u_2}-\varphi_{u_1})].
\end{equation}
 This atomic tunneling  current  is  a pure sinusoidally alternating  atomic   current with 
 the period $T=\pi/2g$.  without $dc$ component. Hence no Shapiro-like steps appears. 

It is interesting to note that when the interatomic nonlinear interactions
are involved, but $q=\chi\neq 0$, we get the same results with those of
the case of $q=\chi=0$ for the atomic tunneling  current and the Shapiro-like steps.
This indicates that the contributions of the nonlinear self-interaction ($q$)
in each condensate to the atomic tunneling  can counteract that of the interspecies nonlinear
interactions ($\chi$). It is the competition between the nonlinear self-interaction and
the interspecies nonlinear interaction that leads to a simple form of the atomic tunneling  
current and the suppression of the Shapiro-like steps.

In particular, we note that when $g=0$, i.e., there is no tunneling
coupling,  and $q\neq \chi$, we find that 
\begin{eqnarray}
\hspace{-0.5cm}
I(t)&=&-2|u_1||u_2|(q-\chi)[|u_1|^2\sin((q-\chi)t-\theta(t))  \nonumber \\
& &+|u_2|^2\sin((q-\chi)t+\theta(t))],
\end{eqnarray}
where $\theta(t)$ is given by  Eq.(32). 
In FIG. 4, we display the time evolution of  the atomic tunneling current  (64) 
 when  the tunneling coupling vanishes.   Results  are shown for the case of $q-\chi=0.1$
 when the two condensates are in the initial  state with   $\alpha_1=10$ and $\alpha_2=5$. From
 FIG.4  we see that the evolution of the atomic tunneling current exhibits the CR phenomenon.

 Eq.(64)  implies that the atomic tunneling  current is nonzero, but no the Shapiro-like
 steps appears. From Eq.(64)  we can see  that even though there is no the
   tunneling coupling between the two condensates  $(g=0)$, the atomic tunneling
 between the two condensates  may  happen. This atomic tunneling  is completely induced by the
 nonlinearity of  interatomic interactions which are characterized by  interatomic collisions 
 ($q$ and $\chi$).   Therefore,  we may conclude  that the nonlinearity of  interatomic interactions 
 in the two  condensates can lead to the atomic tunneling between the two condensates.

\section{Concluding Remarks}
 We have studied the MQST and the quantum coherent atomic tunneling  in a two species
Bose  condensate system in the presence of nonlinear
self-interaction of each species, the interspecies nonlinear interaction, and the
Josephson-like tunneling interaction, and have given new insight
to the MQST and the atomic tunneling. We have shown that the interatomic nonlinear
interactions in the two condensates  induce not only the MQST 
but also the CR of the population difference between two condensates, 
The CR phenomenon can be considered as a novel macroscopic quantum effect. We have indicated that 
the nonlinear interactions significantly affect   the atomic tunneling, and the Shapiro-like steps 
of the atomic tunneling  current. Comparing with Smerzi and coworkers' work [37], 
the present work involves the interspecies nonlinear  interaction.  The involvement of the interspecies 
nonlinear interaction gives rise to new characteristics on 
the MQST and the atomic tunneling. We have shown that when both the nonlinear self-interaction $(q)$ and 
the interspecies nonlinear interaction $(\chi)$ present at the same time, 
the atomic tunneling dynamics and the MQST and the Shapiro-like steps depends upon 
 the difference $(q-\chi)$, not $q$ and $\chi$ themselves.
 We have also found   that the interspecies nonlinear interaction generates the MQST  
at the same level   with nonlinear self-interaction.  However, contribution from the interspecies  interaction  
to the MQST and  that from the self-interaction weaken themselves with each other. 
 It is  the competing effects  between the nonlinear self-interaction in each species and the interspecies
nonlinear interaction that  leads to the quenching of the MQST and the suppression
  of the CR and the Shapiro-like steps of the atomic tunneling  current. Especially, we have revealed that the
nonlinearity of  interatomic interactions in the two condensates  can induce
the coherent atomic tunneling  between two condensates occurs, even though there does not 
exist the Josephson-like tunneling coupling. It should be mentioned that these results are obtained
under  the two-mode approximation and the rotating wave approximation,
so they are valid for  weak nonlinear interactions between atoms.
Finally, It should be noted that in order to observe macroscopic quantum phenomena such as MQST and Shapiro-like steps  
in Bose condensates,  one has to control various interactions between atoms. This controlling can be carried  
out through manipulating  interatomic scattering lengthes. Several  theoretical and experimental approaches [46-51] 
to alter the scattering length have been  proposed. In particular, recent experiments on Feshbach resonances 
in a Bose condensate [46,47]   have indicated  that the scattering length of ultracold atoms can be altered through 
Feshbach resonance. These experimental progresses provide the possibility to observe the MQST and Shapiro-like steps 
in Bose condensates.

\begin{flushleft}
{\large \bf
Acknowledgments}
\end{flushleft}
L. M. Kuang would like to  acknowledge the  Abdus Salam
International Center for Theoretical Physics, Trieste,
for its hospitality where part of this work was done.
This work was supported in part by  the climbing project of China, NSF of China, 
 Educational Committee Foundation and   NSF of Hunan Province,  special project 
 of NSF of China via Institute of theoretical  Physics, Academia Sinica.

\begin{center}
{\bf Figure Captions}
\end{center}

FIG. 1.  Diagram of  the time evolution of  the population  difference   between the two
condensates. The time is in units of   $|q-\chi|$. Results  are shown for  
the case of   $g=25|q-\chi|$, when the two condensates are in the initial  state with 
 $\alpha_1=5$ and $\alpha_2=4$.

FIG.2. Diagram of the time  evolution of the fractional population imbalance. 
The time is in units of  $|q-\chi|$.  Results  are shown for  
(a) $K=1$, and (b) $K=20$ when the two condensates are in the initial  state with 
 $\alpha_1= 10 $ and $\alpha_2=0$.

FIG.3.  The atomic tunneling current between the two
condensates as a function of time $t$ (in arbitrary units).   Results  are shown for the case of 
 $g=0.25$ and $q-\chi=0.1$  when the two condensates are in the initial  state with  
$\alpha_1=5$ and $\alpha_2=4$.

FIG.4.   The atomic tunneling current between the two condensates  as a function of time 
(in arbitrary units) when there does not exist  the tunneling coupling.  Results  are shown for  
 the case of $q-\chi=0.1$  when the two condensates   are in the initial  state with  
  $\alpha_1=10$ and $\alpha_2=5$.

\end{multicols}

\begin{references}
\bibitem.  C. K. Law, H. Pu, N. P. Bigelow, and J. H. Eberly,  \ {\em Phys. Rev. Lett.}
           {\bf 79}, 3105 (1997); H. Pu and N. P. Bigelow,  \ {\em Phys. Rev. Lett.} {\bf 80}, 1130 (1998); 
           {\it ibid}  {\bf 80}, 1134 (1998).
\bibitem.  T. Busch, J. I. Cirac, V. M. Perez-Garcia and P. Zoller, \ {\em Phys. Rev.}  {\bf A56}, 2978 (1997).
\bibitem.  M. R. D. S. Hall, D. S. Jin, J. R. Ensher, C. E. Wieman, and E. A. Cornell, 
          \ {\em Phys. Rev. Lett.} {\bf 81}, 243 (1998); 
           H.-J. Miesner, D. M. Stamper-Kurn, J. Stenger, S. Inouye, A. P. Chikkatur, and W. Ketterle, 
          \ {\em Phys. Rev. Lett.} {\bf 82}, 2228 (1999).
          
\bibitem.   R. Graham and D. Walls,  \ {\em Phys. Rev.} {\bf A57}, 484 (1998).
\bibitem.   B. D. Esry and C. H. Greene,  \ {\em Phys. Rev.} {\bf A57}, 1265 (1998);
            B. D. Esry, {\it et al.} ,  \ {\em Phys. Rev. Lett.} {\bf 78}, 3594 (1997).
\bibitem.  C. J. Myatt {\it et al.}, \ {\em Phys. Rev. Lett.} {\bf 78}, 586 (1997).
\bibitem.  D. S. Holl, M. R. Matthews, C. E. Wieman, and E. A. Cornell,  \ {\em Phys. Rev. Lett.} {\bf 81}, 1543 (1998).
\bibitem.  D. S. Holl, M. R. Matthews, J. R. Ensher, C. E. Wieman, and E. A. Cornell,  \ {\em Phys. Rev. Lett.} 
           {\bf 81}, 1539 (1998).
\bibitem.  T.-L. Ho and V. B. Shenoy,  \ {\em Phys. Rev. Lett.} {\bf 77}, 3276 (1996).
\bibitem.  A. Sinatra, P. O. Fedichev, Y. Castin, J. Dalibard, and G. V. Shlyapnikov, 
           \ {\em Phys. Rev. Lett.} {\bf 82}, 251 (1999).
\bibitem.   E. Tmmermans,  \ {\em Phys. Rev. Lett.} {\bf 81}, 5718 (1998).
\bibitem.  C. K. Law, H. Pu, N. P. Bigelow, and J. H. Eberly,  \ {\em Phys. Rev.} {\bf A58}, 531 (1998).
\bibitem.  B. D. Esry, and C. H. Greene,  \ {\em Phys. Rev.} {\bf A59}, 1457 (1998);
\bibitem.   J. Ruostekoski and D. F. Walls,  \ {\em Phys. Rev.} {\bf A58}, R50 (1998).
\bibitem.   R. W. Spekkens and J. E. Sipe,  \ {\em Phys. Rev.} {\bf A59}, 3868 (1999).
\bibitem.   J. F. Corney, G. J. Milburn, and Weiping Zhang,  \ {\em Phys. Rev.} {\bf A59}, 4630 (1999).
\bibitem.   N. Tsukada, M. Gotoda, Y. Nomura, and T. Isu,  \ {\em Phys. Rev.} {\bf A59}, 3862 (1999).
\bibitem.  M. R. Andrews, {\it et al.} ,  \ {\em Science} {\bf 275}, 637 (1997).
 \bibitem. R. J. Ballagh, K. Burnett and T. F. Scott,  \ {\em Phys. Rev. Lett.} {\bf 78}, 1607 (1997).
\bibitem.  M. -O. Mewes, {\it et al.} ,  \ {\em Phys. Rev. lett.} {\bf 78}, 582 (1997).
\bibitem.  J. Javanainen and S. M. Yoo, {\em Phys. Rev. Lett.} {\bf 76}, 161 (1996);
           J. Javanainen and M. Wilkens, {\em Phys. Rev. Lett.} {\bf 78}, 4675 (1997).
\bibitem.  M. J.Steel and D. F. Walls, {\em Phys. Rev.} {\bf A57}, 3805 (1998).
\bibitem.  T. Wong, M. J. Collett, and D. F. Walls,  \ {\em Phys. Rev.} {\bf A54}, R3718 (1996);
\bibitem.  M. Naraschewski, H. Wallis, and A. Schenzle,  \ {\em Phys. Rev.} {\bf A54}, 2185(1996);
           J. I. Cirac, C. W. Gardiner, M. Naraschewski, and P. Zoller, \ {\em Phys. Rev.} {\bf A54}, R3714 (1996);
            H. Wallis, A. R\"{o}hrl, M. Naraschewski,  and A. Schenzle,    \ {\em Phys. Rev.} {\bf A55}, 2109 (1997).
\bibitem.  W. Hoston and L. You, \ {\em Phys. Rev.} {\bf A53}, 4254 (1996);
\bibitem.  J. Javanainen and M. Yoo,  \ {\em Phys. Rev. Lett.} {\bf 76}, 161 (1996);
\bibitem.  I. Zapata, F. Sols, and A. J. Leggett,
           \ {\em Phys. Rev.} {\bf A57}, R28 (1998).
\bibitem.  J. Javanainen, {\em Phys. Rev. Lett.} {\bf 57}, 3164 (1986).
\bibitem.  S. Grossmann and M. Holthaus,  \ {\em Z. Naturforsch} {\bf 50a}, 323 (1995).
\bibitem. G. J. Milburn, J. Corney, E. M. Wright, and D. F. Walls, \ {\em Phys. Rev.}  {\bf A55}, 4318 (1997).
\bibitem.  J. I. Cirac, M. Lewenstein, K. Molmer, and P. Zoller, \ {\em Phys. Rev.}  {\bf A57}, 1208 (1998).
\bibitem.  M. J. Steel and M. J. Collett, \ {\em Phys. Rev.}  {\bf A57}, 2920 (1998).
\bibitem.   J. Williams, R. Walser, J. Cooper, E. Cornell, and M. Holland, \ {\em Phys. Rev.}  {\bf A59}, R31 (1999).
\bibitem. Patrik \"{O}hberg and Stig Stenholm, \ {\em Phys. Rev.}  {\bf A59}, 3890 (1999).
\bibitem.  E. M. Wright, D. F. Walls, and J. C. Garrison, {\em Phys. Rev. Lett.} {\bf 77}, 2158 (1996);
           E. M. Wright, T. Wong, M. J. Collett, S. M. Tan, and D. F. Walls,
           {\em Phys. Rev.} {\bf A56}, 591 (1997).
\bibitem.  Le-Man Kuang and Zhao-Yang Zeng, {\em  Chin. Phys. Lett.} {\bf 15}, 703 (1998).
\bibitem.  A. Smerzi, S. Fantoni, S. Giovanazzi and S. R. Shenoy,
          \ {\em Phys. Rev. Lett.} {\bf 79}, 4950 (1997);
           S. Raghavan, A. Smerzi, S. Fantoni, and S. R. Shenoy,
           \ {\em Phys. Rev.} {\bf A59}, 620(1999).
\bibitem.  {\it Single Charge Tunneling, Coulomb Blockade Phenomena in Nanostructures}, edited by 
           H. Grabert and M. H. Devoret (Plenum, New York, 1991).
\bibitem.  V. M. Kenkre and D. K. Campbell, \ {\em Phys. Rev.} {\bf B34}, 4959 (1986).
\bibitem.   P. W. Anderson, \ {\em Rev. Mod. Phys.} {\bf 38}, 298 (1966).
\bibitem.   O. Avenel and E. Varoquaux, \ {\em Phys. Rev. Lett.} {\bf 55}, 2704 (1985).
\bibitem.  N. Korolkova, J. Perina, \ {\em  Opt. Commun.} {\bf 136}, 135 (1996).
\bibitem.  A. P. Alodjanc, S. M. Arakeljan, and A. S. Chirkin,   \ {\em JETP} {\bf 81}, 34 (1995).
\bibitem.  J. H. Eberly, N. B. Narozhny, and J. J. Sanchez-Mondragon,  {\em Phys. Rev. Lett.} {\bf 44},
           1323 (1980).
\bibitem.  L. Solymar, \ {\em  Superconductive Tunneling and Applications} (Chapman and Hall, London, 1972);
            A. Barone and G.  Paterno,  \ {\em  Physics and  Applications of the Josephson Effect} (Wiley, New York, 1982).
\bibitem.   S. Inouye, M. R. Andrews, J. Stenger, H.-J. Miesner, D. M. Stamper-Kurn, and W. Ketterle, 
           \ {\em  Nature} {\bf 392}, 151 (1998).
\bibitem.   J. Stenger, S. Inouye, M. R. Andrews, H.-J. Miesner, D. M. Stamper-Kurn, and W. Ketterle, 
            \ {\em  Phys. Rev. Lett.} {\bf 82}, 2422(1999).
\bibitem.   Ph. Courteille, R. S. Freeland, and D. J. Heinzen, F. A. van Abeelen, and B. J. Verhaar, 
           \ {\em  Phys. Rev. Lett.} {\bf 81}, 69 (1998).
\bibitem.   J. M. Vogels, C. C. Tsai, R. S. Freeland, S. J. J. M. F. Kokkelmans, B. J. Verhaar, and D. J. Heinzen, 
           \ {\em  Phys. Rev.} {\bf A56}, R1067 (1997);
            E. Tiesinga, B. J. Verhaar, and H. T. C. Stoof, \ {\em Phys. Rev.} {\bf A47}, 4114 (1993);
            A. J. Moerdijk, B. J. Verhaar, and A. Axelsson, \ {\em Phys. Rev.} {\bf A51}, 4852 (1995).
\bibitem.   J. L. Bohn and  P. S. Julienne, \ {\em Phys. Rev.} {\bf A56}, 1486 (1997).
\bibitem.   Yu. Kagan, E. L. Surkov, and G. V. Shlyapnikov, \ {\em Phys. Rev. Lett.} {\bf 79}, 2604 (1997);
            P. O. Fedichev, Yu. Kugan, G. V. Shlyapnikov, and J. T. M. Walraven, \ {\em Phys. Rev. Lett.} {\bf 77}, 
             2913 (1996).
\end{references}
\end{document}